\documentclass[a4paper,twoside,10pt]{article}
\usepackage{graphicx,publaob}
\usepackage{epstopdf}

\def\papertype{Progress report}
\def\sun{\odot}
\def\deg{$^{\rm o}$}
\begin{document}

% Title
\title{REGULUS OBSERVED WITH VLTI/AMBER}

% Authors
\authors{S. JANKOV$^1$, M. HADJARA$^{2,3,4}$ ,  R. G. PETROV$^4$, \linebreak \newline
 P. CRUZALEBES$^4$, A. SPANG$^4$ \lowercase{and} S. LAGARDE$^4$}

% Addresses and e-mails
\address{$^1$Astronomical Observatory, Volgina 7, 11000 Belgrade, Serbia}
\Email{sjankov}{aob}{rs}
\address{$^2$Instituto de Astronom\`{\i}a, Universidad Cat\`olica del Norte, Av. Angamos 0610 Antofagasta, Chile}
%\Email{massinissa.hadjara}{gmail}{com}
\address{$^3$Centre de Recherche en Astronomie, Astrophysique et G\' eophysique (CRAAG), 
Route de l'Observatoire, B.P. 63, Bouzareah, 16340, Alger, Alg\' erie}
%\Email{massinissa.hadjara}{gmail}{com}
\address{$^4$Universit\' e  de Nice-Sophia Antipolis,  Campus Valrose, 06108 Nice cedex 2, France}
%\Email{Romain.Petrov}{unice}{fr}

% Running titles
\markboth{REGULUS OBSERVED WITH VLTI/AMBER}{S. JANKOV, M. HADJARA and R. G. PETROV}

% Abstract
\abstract{
The rapidly rotating primary component of Regulus A system has been observed, 
for the first time, using the technique of differential interferometry at high 
spectral resolution. The observations have been performed across the Br$_\gamma$ spectral line with the VLTI/AMBER focal instrument in high spectral resolution mode  (R $\approx 12\,000$) at $\approx 80-130$m (projected on the sky) Auxiliary Telescopes triplet baseline configurations. We confirm, within the uncertainties, the results previously obtained using the techniques of classical long-baseline interferometry, although the question of anomalous gravity darkening remains open for the future study.
}

% Section and subsection
\section{INTRODUCTION}

Located at a distance of $23.76 \pm 0.04$ pc (van Belle \& von Braun 2009), $\alpha$ Leo is a multiple stellar system composed of at least two binaries. The A component of the system ($\alpha$ Leo A, HD 87901) has been recently discovered to be a spectroscopic binary where the fainter companion is probably a white dwarf or a M4 V star of mass $\approx 0.3 $M$_\sun$ and an orbital period of 40.11 days  (Gies et al. 2008) . The brighter companion (hereafter Regulus) was classified as a main sequence B7V star by Johnson \&  Morgan (1953), and more recently as a sub-geant B8IV star by Gray et al. (2003) of mass $\approx 4 $ M$_\sun$  (Che et al. 2011, and references therein).

It is well known (by observing even with an amateur telescope) that $\alpha$ Leo A has another companion which is in fact  a system of two other components (B and C) which together form a binary system (McAlister et al. 2005).  This B-C subsystem is located $\approx$ 3$'$ from the A component. 
The B component ($\alpha$ Leo B; HD 87884) is a $\sim$ 0.8 M$_\sun$ star of spectral type K2 V, while  the C component is a very faint M4 V star with a mass of $\sim$ 0.2 M$_\sun$. The Washington Double Star Catalog (Mason et al. 2001)  lists a D component, also having a common proper motion with the system and a separation of $\approx$  3.6$'$ from the A component.  At such separations the B-C subsystem and D component have never directly interacted with Regulus, but the fainter  component of the $\alpha$ Leo A subsystem influences profoundly its evolution through mass exchange between two stars (Rappaport, Podsiadlowski, \& Horev 2009). Considering a scenario where the initial masses of the progenitors of fainter component and Regulus are 2.3 $\pm$0.2 M$_\sun$ and 1.7 $\pm$  0.2 M$_\sun$ respectively, they infer  the age of the system which exceeds 1 Gyr. They also consider a possibility that the mass transfer is the cause of the current rapid rotation of Regulus.

Regulus has been identified as a fast rotator by Slettebak (1954), who determined spectroscopically its high rotational velocity Veq $\sin i$ = 352  $\pm$ 8 km/s. The first interferometric observations of the star weredone with the Narrabri Intensity Interferometer by Hanbury Brown et al. (1974), but only the information about its size could be obtained. They derived the equatorial angular diameter 1.32  $\pm$ 0.06 mas.
Such a diamater, together with the high apparent brightness of Regulus made it a very good interferometric target allowing to reveal its extremely oblate shape (McAlister et al. 2005), and to confirm the spectacular discovery of the extremely oblateness of Achernar (Domiciano et al. 2003). A very important consequence of stellar oblateness is the associated gravity darkening (von Zeipel 1924a,b) which implies a variation of associated effective temperature over the stellar surface.  
For Regulus McAlister et al. (2005) determined a difference of $\sim$ 5000 K between the poles and the equator, a finding which has been confirmed by Che et al. (2011). Such a large difference in associated effective surface temperatures make the spectral classification quite a challenging task, and should be taken into account in the modeling of observations as well as in theoretical analysis concerning the evolutionary status of the entire binary system.

By the other hand Regulus is very challenging object for investigation because of its almost equator-on orientation which makes the situation where the minimization procedures can produce the degenerated solutions. For this reason we observed the star, for the first time, with the VLTI/AMBER instrument which provides the differential interferometry data, in order to check the results previously published using the classical interferometry instruments. 

\section{OBSERVATIONS}
%\subsection{OBSERVATIONS}

Regulus was observed with the AMBER an interferometric near infrared focal instrument for the Very Large Telescope Interferometer (VLTI), using the Auxiliary Telescopes (AT).
The observations have been performed in the high spectral resolution mode of AMBER ($\approx$ 12000), centered on the Br$_{\gamma}$ spectral line. The corresponding $u,v$ (Fourier space) coverage is shown in Fig.1, while the Table 1 provides our observing log.

The interferometric fringes were stabilized using a fringe tracker FINITO (M\'erand et al. 2012). 
The instrument AMBER provides both the intensity spectrum and the relative phase of the interferometric signal, and it is described in detail by Petrov et al. (2007), Robbe-Dubois et al. (2007) and in references therein.

Thye intensity spectrum is the {\it zero order moment} of the sky brightness distribution while the photocenter shift spectrum is the {\it first order moment} of the sky brightness distribution (Jankov et al. 2001, and references therein). {\it It is a vectorial function} and can be evaluated measuring the relative phase of the interferometric signal along a spectral line with respect to the continuum. In order to study the physics of a star the photocenter shift measurements should be evaluated in the coordinate system related to the stellar  rotational axis (${\epsilon}_{\rm eq}$ \& ${\epsilon}_{\rm p}$), equatorial and polar components respectively.
 
However, {\it a priori} only the components related to the celestial coordinate system (${\epsilon}_{\alpha}$ \& ${\epsilon}_{\alpha}$) can be evaluated. Figure 2. shows these components as observed in our observing run.
 
In order to evaluate the components related to the stellar rotational axis, the position angle of the axis (PArot) should be determined. It can be done in the global procedure for determination of stellar parameters as well as independently based on the method described by Petrov \& Lagarde (1992), as shown in the Fig. 3. Then the equatorial and polar components of the photocenter shift (see Fig. 4) can be obtained by simple rotation of the coordinate system.
The Fig 5. shows the 3D representation of the observed photocenter shift components of Regulus.

% Figure (in PS or EPS format)
\begin{figure}
\includegraphics[width=12cm,height=8cm]{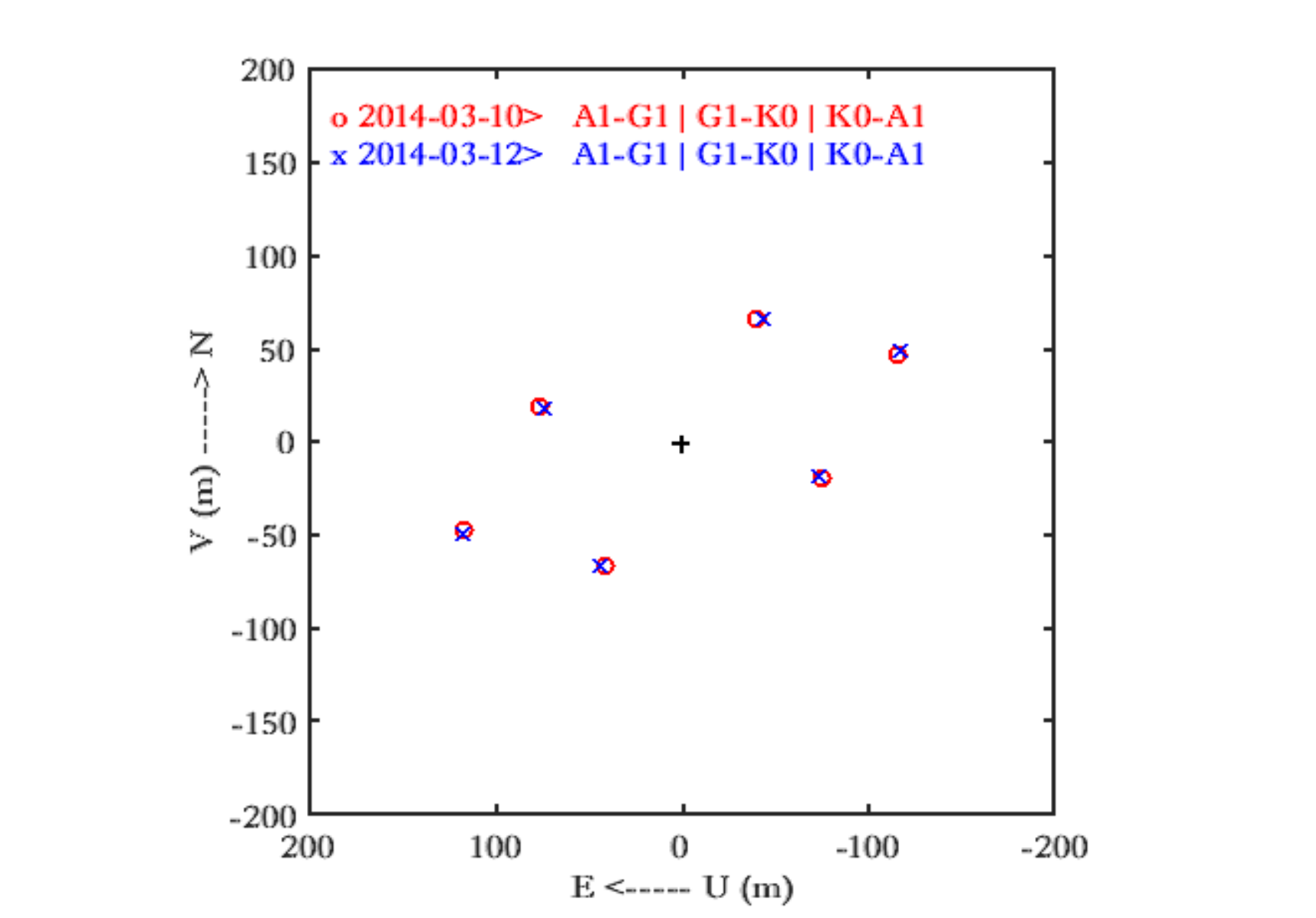}
\caption{ (u,v) coverage of VLTI/AMBER observations of Regulus; spanned over 1.5 h/night. Note the rather poor sampling of the Fourier space. According to the Table 1, the $(u,v)$ points for the date 2014-03-10 are represented by circles and by crosses for the date 2014-03-12.}
\end{figure}

%\vfil\newpage

% Table
\begin{table}
\begin{center}
\begin{tabular}{|c  c c c|}
\hline
   Object    & Date and time & Baseline (m) & Baseline PA (\deg) \\
\hline
60 Cnc & 2014-03-10T03:09 & 75,81,128 & 103,34,67 \\
Regulus & 2014-03-10T03:48 & 78,77,125 & 104,32,68 \\
w Cen & 2014-03-10T04:30 & 73,87,129 & 88,14,47 \\
w Cen & 2014-03-10T04:44 & 74,87,129 & 90,16,50 \\
\hline
$\epsilon$  Cma & 2014-03-12T02:15 & 75,87,116 & 124,36,76 \\
Regulus & 2014-03-12T03:59 & 76,79,127 & 104,33,68 \\
w Cen & 2014-03-12T04:45 & 75,87,129 & 92,17,51 \\
$\iota$ Cen & 2014-03-12T07:17 & 80,88,126 & 113,30,69 \\
\hline
\end{tabular}
\caption{VLTI/AMBER observations of Regulus and its calibration stars using the AT triplet A1-G1-K0.}
\end{center}
\end{table}

% Figure (in PS or EPS format)
\begin{figure}
\includegraphics[width=12cm,height=8cm]{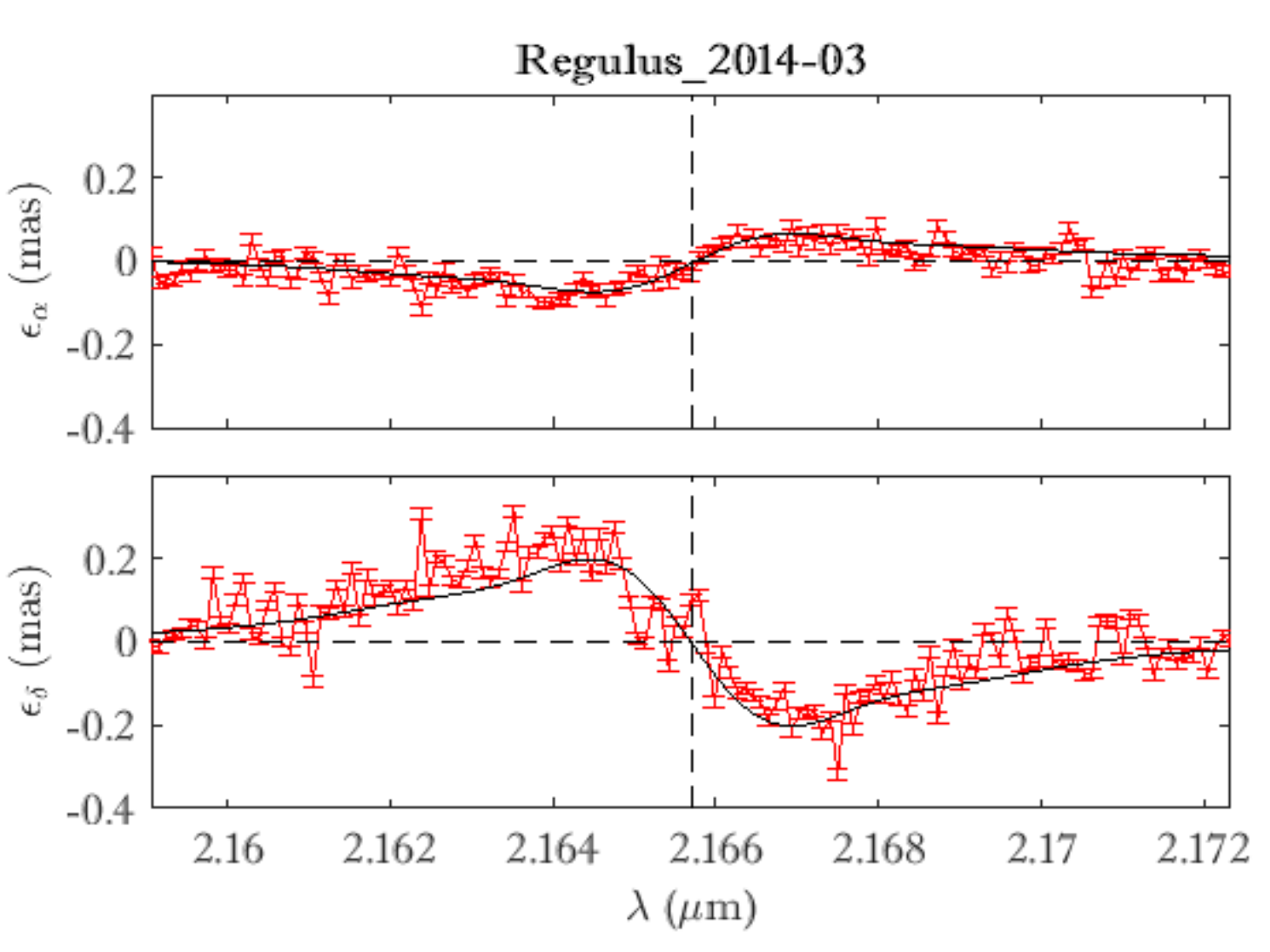}
\caption{The spectra of perpendicular photocenter components ${\epsilon}_{\alpha}$ \& ${\epsilon}_{\delta}$ }
\end{figure}

% Figure (in PS or EPS format)
\begin{figure}
\includegraphics[width=12cm,height=8cm]{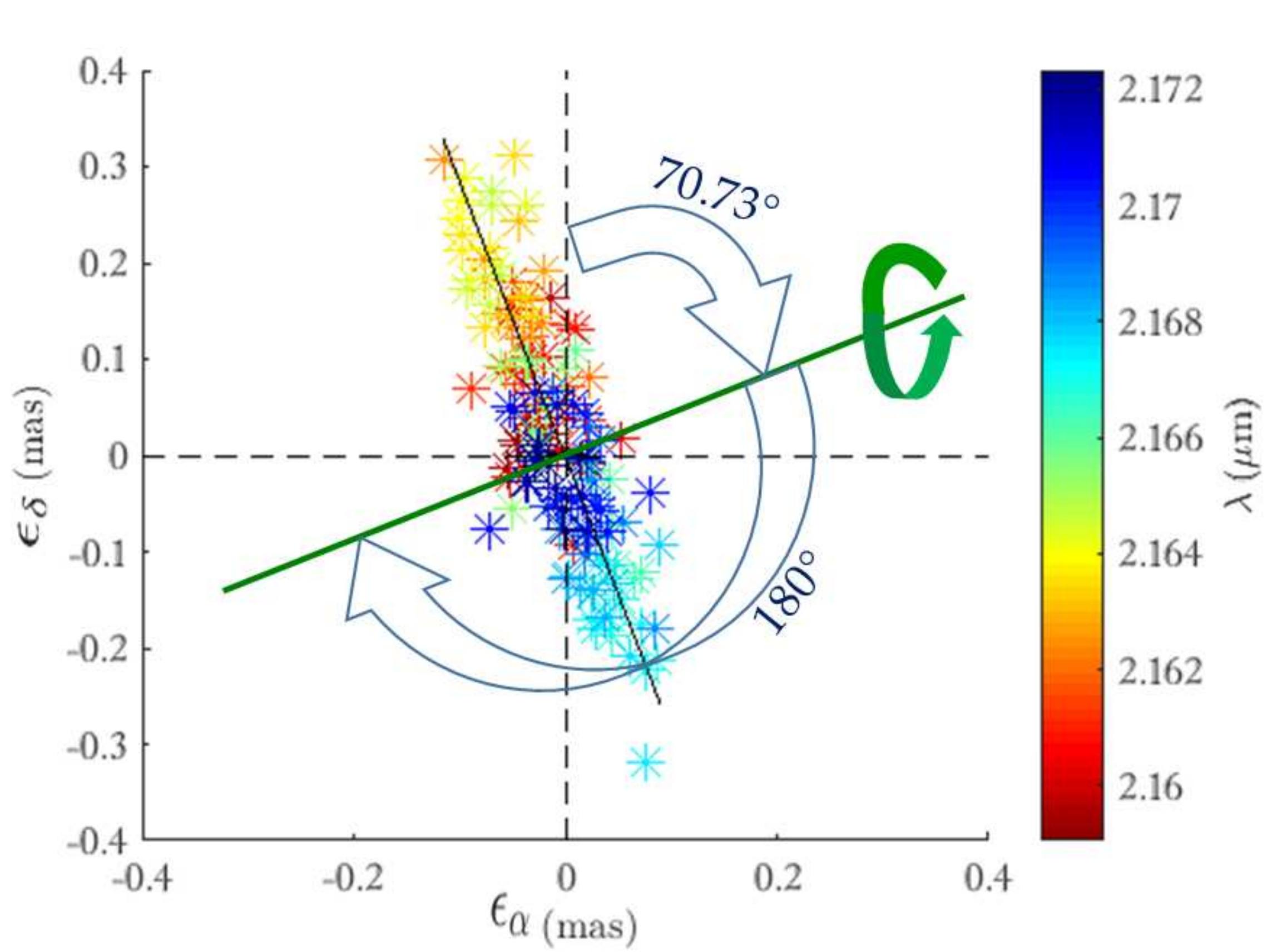}
\caption{ The position angle of the rotational axis from observed photocenter shifts}
\end{figure}

% Figure (in PS or EPS format)
\begin{figure}
\includegraphics[width=12cm,height=8cm]{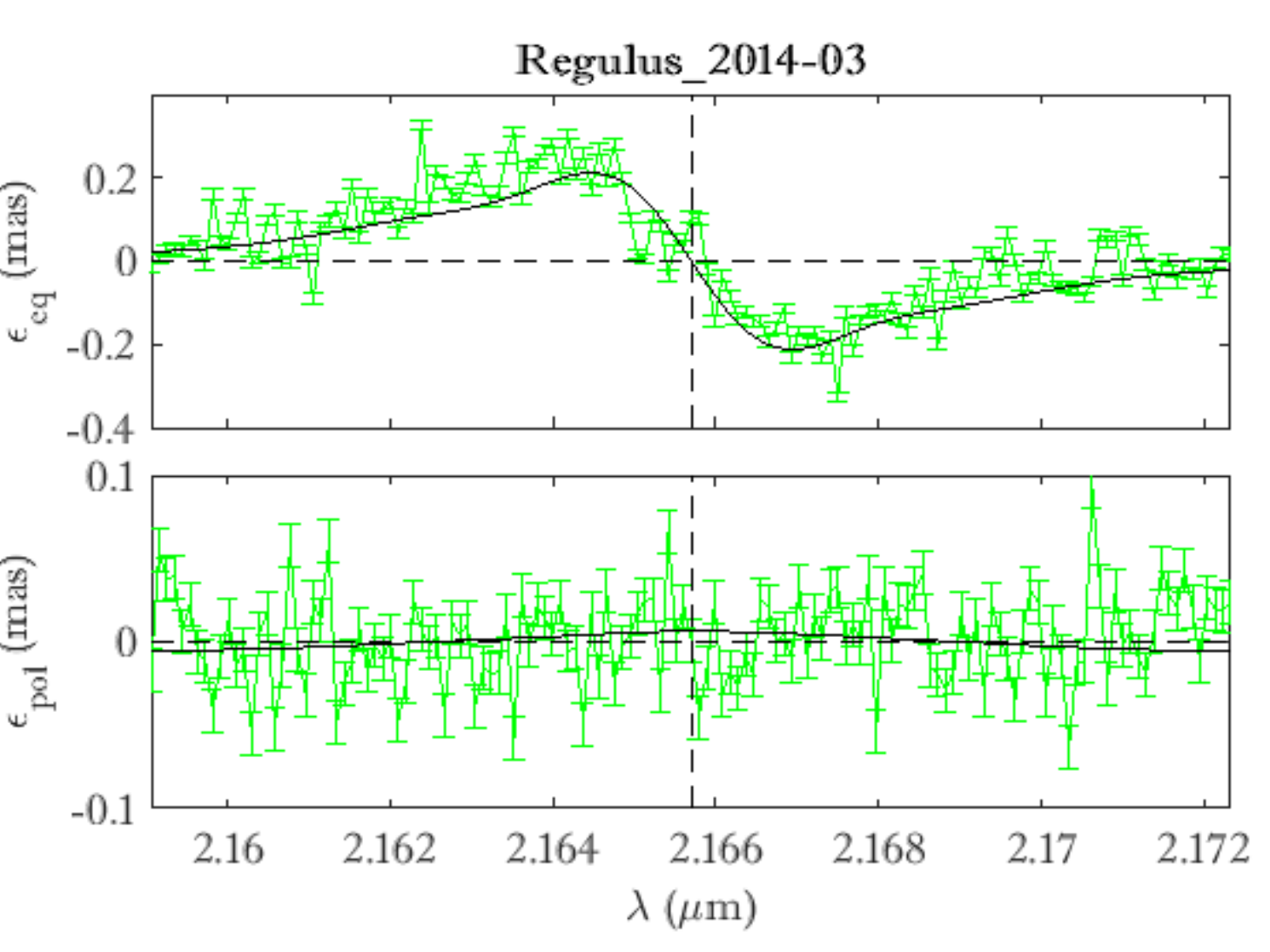}
\caption{The spectra perpendicular photocenter components ${\epsilon}_{\rm eq}$ \& ${\epsilon}_{\rm p}$}
\end{figure}

% Figure (in PS or EPS format)
\begin{figure}
\includegraphics[width=12cm,height=8cm]{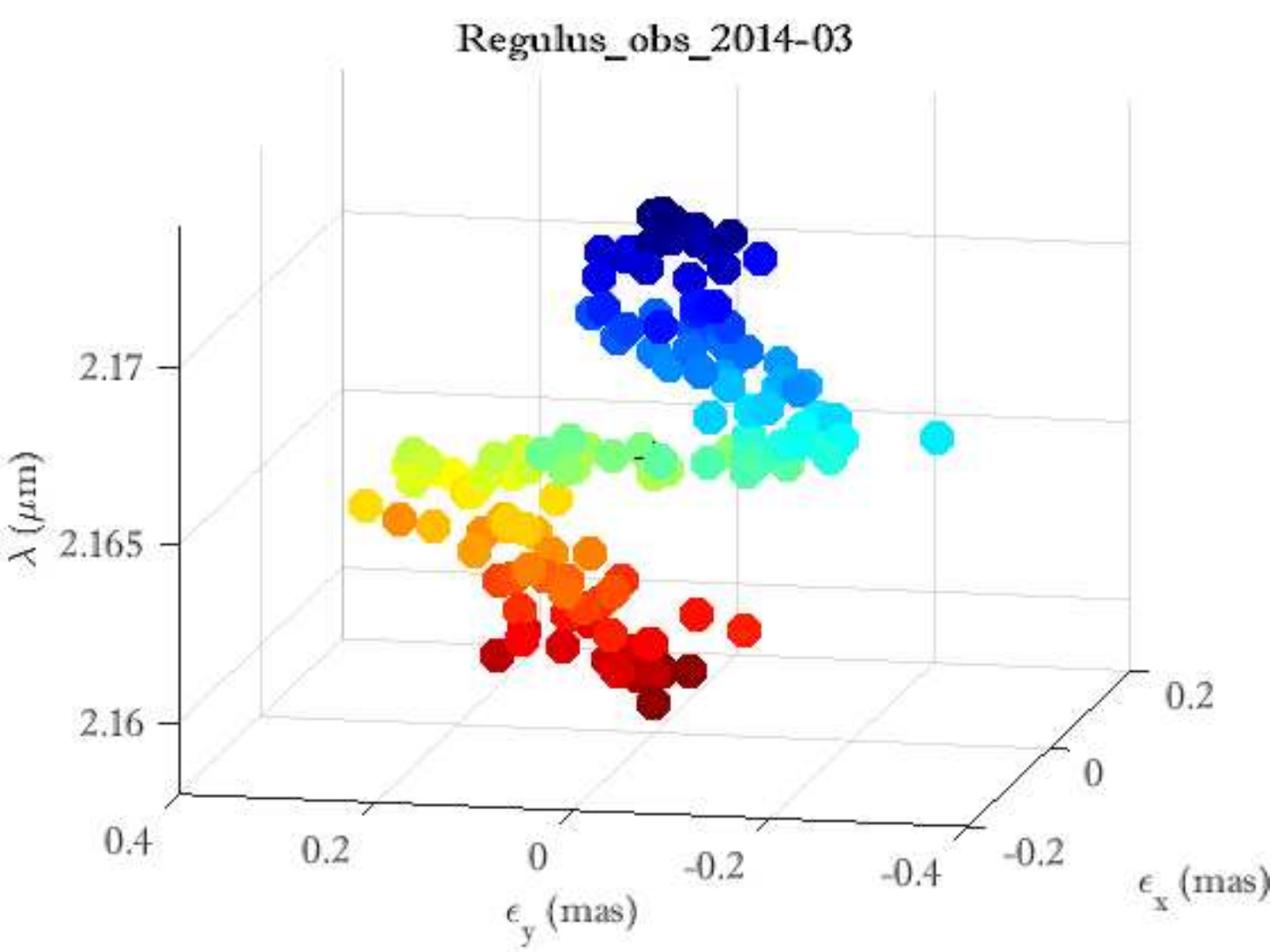}
\caption{ Observed photocenter shifts of Regulus}
\end{figure}

\section{RESULTS}

To interpret the observations, we use the semi-analytical model for fast rotators:  Simulation Code of Interferometric observations for ROtators and CirCumstellar Objects (SCIROCCO). This code written in Matlab, allows to compute monochromatic intensity maps of uniformly rotating, flattened, and gravity darkened stars using the semi-analytical approach. The code SCIROCCO, which is a parametric description of the velocity field and the intensity map for line profile modelisation, is described in detail in Hadjara et al. (2014); Hadjara (2015).

In order to deduce the best parameters we perform a $\chi^2$ minimization, and the corresponding modeled photocenter shifts are shown in the  Fig. 6.
To check whether the correct global minimum is achieved, in addition to the non-stochastic $\chi^2$ minimization method, we use a stochastic Markov Chain Monte Carlo (MCMC) method as well. The results are summarized and compared to the previously published results in the Table 2.

% Figure (in PS or EPS format)
\begin{figure}
\includegraphics[width=12cm,height=8cm]{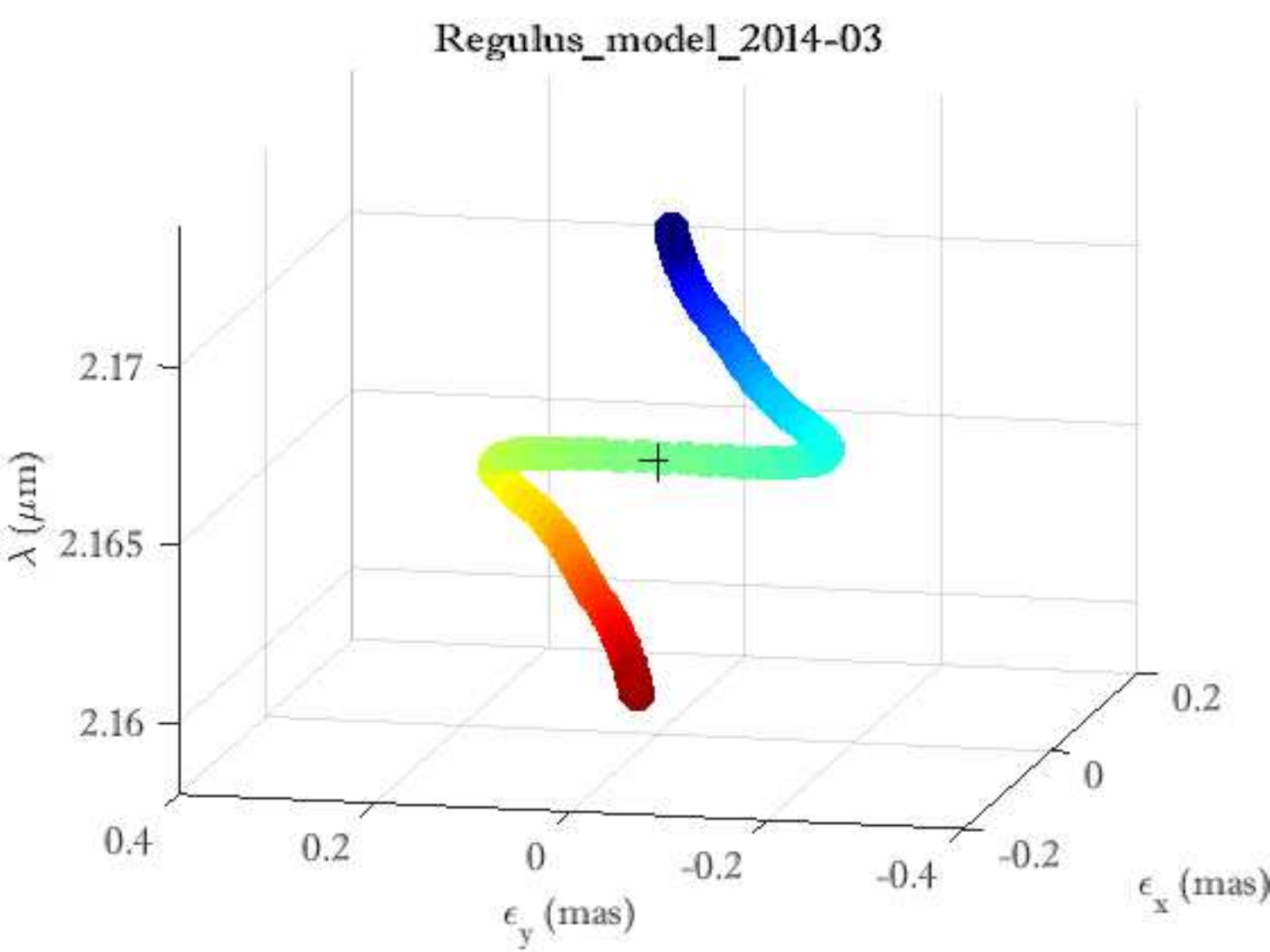}
\caption{ Modeled photocenter shifts of Regulus}
\end{figure}

\begin{table}
\begin{center}
\begin{tabular}{|c|  c c c c|}
\hline
   &$\chi^2$ & MCMC & Mc Alister et al. (2005) & Che et al. (2011)\\
\hline
 &  &  &  & \\
Req [R$_\sun$]& $4.2\pm 0.1$ & $4.2\pm 0.3$ & $4.16\pm 0.08$ & $4.21_{-0.06}^{+0.07}$\\
Veq $\sin i$ [km/s]& $350 \pm 15$ & $360\pm 30$ & $317_{-85}^{+3}$ & $340_{-30}^{+20}$\\
PArot [\deg]& $251\pm 4$ & $253\pm 2$ & $266\pm 3$ & $258_{-1}^{+2}$\\
$i$ & $86\pm 2$ & $86\pm 10$ & $90_{-15}^{+0}$ & $86_{-2}^{+1}$\\
Teq [K] & $10600 \pm 600$& $10600 \pm 800$& $10300 \pm 1000$& $11000 _{-500}^{+400}$\\
Tp [K] & $144000 \pm 800$& $14400 \pm 1100$& $154000 \pm 1400$& $14500 _{-700}^{+600}$\\
\hline
\end{tabular}
\caption{The parameters deduced with $\chi^2$ \& MCMC minimization as well as previously published data}
\end{center}
\end{table}

We complete our study by examining the coupling of gravity darkening coefficient and inclination ($\beta , i$). We deduce the probability space that shows the degeneracy of corresponding stellar parameters in the Fig. 7, where we superimposed the Regulus probability space ($\beta , i$) of Che et al. (2011) to ours. We can observe a strong degeneration between the couple. This figure shows a strongly enlarged contour of the probability, implying an important correlation between $\beta$ and $i$, which means that we must not rely only on the $\chi^2$ minima. This subject has beeen studied in more details by Hadjara et al. (2017).

% Figure (in PS or EPS format)
\begin{figure}
\includegraphics[width=12cm,height=8cm]{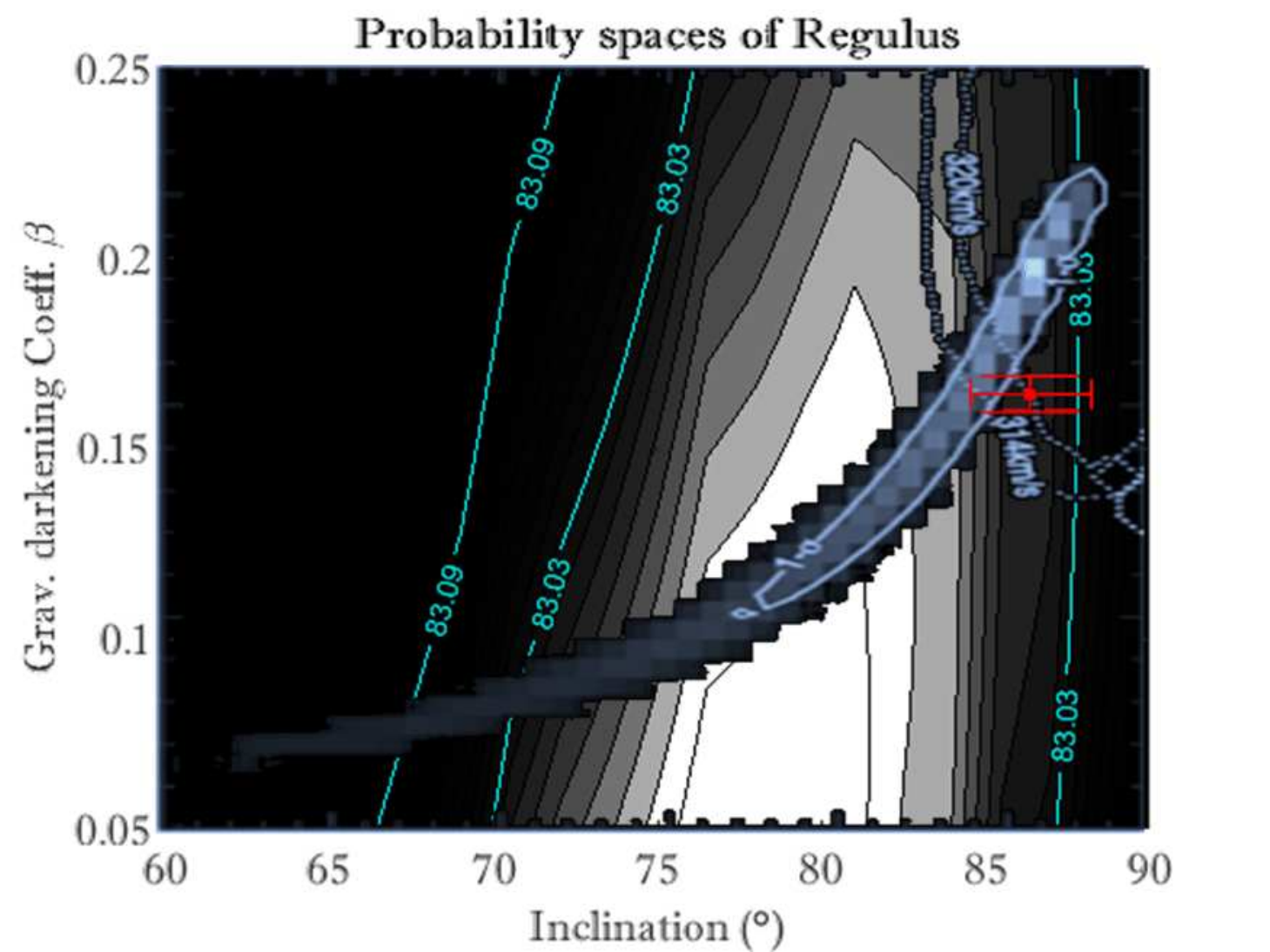}
\caption{Probability space for Regulus that shows the degeneracy between the stellar parameters $\beta$ and $i$ in shades of gray. The solid contours represent the 83\% probability, and the with the error bars is the best model fitting result for $i$ and $\beta$ from Espinoza Lara \& Rieutord (2011).}
\end{figure}

\section{CONCLUSIONS}

Using, for the first time, the approach of spectrally resolved differential interferometry we confirm the results previously obtained for Regulus by long-baseline interferometry (McAlister et al. 2005; Che et al. 2011). However we note that the results obtained for the gravity darkening coefficient should be considered with a great caution which makes this subject open for the future study.

% References
\references

Che X., et al., 2011,  \journal{ApJ}, \vol{732}, 68

Domiciano de Souza A., Kervella P., Jankov S., Abe L., Vakili F., di Folco E., Paresce F., 2003,  \journal{A\&A}, \vol{407}, L47

Gies D. R., et al., 2008,  \journal{ApJ}, \vol{682}, L117

Gray R. O., Corbally C. J., Garrison R. F., McFadden M. T., Robinson P. E., 2003,  \journal{AJ}, \vol{126}, 2048

Hadjara M., Petrov, R. G., Jankov, S. Cruzal\`ebes, P., Spang, A., Lagarde, S., 2017, submitted to \journal{MNRAS} 

Hadjara M., 2015, PhD thesis, Universit\'e de Sophia Antipolis

Hadjara M., Vakili F., Domiciano de Souza A., Millour F., PetrovR., Jankov S., Bendjoya P., 2013, in Mary D., Theys C., Aime
C., eds,  \journal{EAS Publications Series, \vol59}, 131

Hadjara M., et al., 2014, \journal{A\&A}, \vol{569}, A45

Hanbury Brown R., Davis J., Allen L. R., 1974,  \journal{MNRAS}, \vol{167}, 121

Jankov S., Vakili F., Domiciano de Souza Jr. A., Janot-Pacheco E., 2001, \journal{A\&A}, \vol{377}, 721

Johnson H. L., Morgan W. W., 1953,  \journal{ApJ}, \vol{117}, 313

Mason, B. D., Wycoff, G. L., Hartkopf, W. I., Douglass, G. G., \& Worley, C. E. 2001,  \journal{AJ}, \vol{122}, 3466

McAlister H. A., et al., 2005,  \journal{ApJ}, \vol{628}, 439

M\'erand A., Patru F., Berger J.-P., Percheron I., Poupar S., 2012, in Optical and Infrared Interferometry III. p. 84451K
(arXiv:1211.2213), doi:10.1117/12.925450

Petrov R. G., Lagarde S., 1992, in McAlister H. A., Hartkopf W. I., eds,  \journal{Astronomical Society of the Pacific Conference
Series}, \vol{32}, 477

Petrov, R. G.,  Malbet, F.,  Weigelt, G, et al., 2007,  \journal{A\&A}, \vol{464}, 1

Rappaport S., Podsiadlowski P., Horev I., 2009,  \journal{ApJ}, \vol{698}, 666

Robbe-Dubois, S.; Lagarde, S.; Petrov et al. 2007,  \journal{A\&A}, \vol{464}, 13

Slettebak A., 1954,  \journal{ApJ}, \vol{119}, 146

van Belle G. T., von Braun K., 2009,  \journal{ApJ}, \vol{694}, 1085

von Zeipel H., 1924a,  \journal{MNRAS}, \vol{84}, 665

von Zeipel H.: 1924b, \journal{MNRAS}, \vol{84}, 684

\endreferences

\end{document}